\documentstyle[prl,aps,epsf,axodraw,12pt]{revtex}
\tightenlines
\begin{document}
\title{Trace anomalies and chiral Ward identities}
\author{Ji-Feng Yang}
\address{Department of Physics, East China Normal University,
Shanghai 200062, China}
\maketitle
\begin{abstract}
In a simple abelian spinor field theory, the canonical trace
identities for certain axial-vector and axial-scalar operators are
reexamined in dimensional regularization, some disagreements with
previous results are found  and an interesting new phenomenon is
observed and briefly discussed.
\end{abstract}
\vspace{0.5cm}
It is well known that chiral anomaly has direct physical and
topological connections\cite{PCAC} and similarly for trace
anomaly\cite{Cole,crew,PCDC}, such anomalies are often termed as
quantum mechanical violation of classical symmetries,
specifically, via the regularization effects in quantum field
theories. That is, the quantization procedure is incompatible with
such symmetries. In 't Hooft's seminal interpretation, chiral
anomaly has also been shown to arise from the decoupling of heavy
fermions\cite{Hooft}, namely, chiral anomaly is closely related to
dynamical contents. Therefore, chiral and other anomalies have
become the key concern for model construction both in field
theories\cite{AW} and string theories\cite{GS}. Thus anomalies in
canonical relations are very important in field theories and high
energy physics, their appearances are often helpful in deepening
our understanding of the quantum theories.

In this letter, we examine the trace anomalies with an emphasis on
the relation between the trace and the chiral identities for
certain axial operators, as they are important in modern particle
physics, especially in the supersymmetric field
theories\cite{susy}. Specifically, we examine the trace and chiral
relations satisfied by the two- and three- point functions of
operators $j^5_{\mu}\equiv \bar{\psi}\gamma_{\mu}\gamma_5\psi$,
$j^5\equiv2im\bar{\psi}\gamma_5\psi$, $\theta\equiv
m\bar{\psi}\psi,$ and $\sigma\equiv 4m\bar{\psi}\psi $ computed in
dimensional regularization. The non-abelian ones have been
examined in Ref.\cite{PCDC} through partial calculation. Here we
carry out all the one loop calculations which are in fact very
simple and then examine explicitly the relations satisfied by
these quantities.

The objects to be calculated are listed as follows,
\begin{eqnarray}
\label{def1} &&{\Pi}^5_{\mu\nu}(p,-p)\equiv i
{\mathcal{FT}}\{\langle 0|T(j_{5\mu}j_{5\nu})|0\rangle\},\\ &&
\Delta^5_{\mu\nu}(0,p,-p)\equiv{\mathcal{FT}}\{\langle 0|T(\theta
j_{5\mu}j_{5\nu})|0\rangle\};\\
 &&{\Pi}^5_{\ \nu}(p,-p)\equiv i{\mathcal{FT}}\{\langle
0|T(j_{5}j_{5\nu})|0\rangle\},\\ &&\Delta^5_{\
\nu}(0,p,-p)\equiv{\mathcal{FT}}\{\langle 0|T(\theta
j_{5}j_{5\nu})|0\rangle\}; \\
&&{\Pi}^5(p,-p)\equiv i{\mathcal{FT}}\{\langle
0|T(j_{5}j_{5})|0\rangle\},\\
&&\Delta^5(0,p,-p)\equiv{\mathcal{FT}}\{\langle 0|T(\theta
j_{5}j_{5})|0\rangle\};\\
&&\langle\sigma\rangle\equiv
{\mathcal{FT}}\{4m\langle\bar{\psi}\psi\rangle\},\\
&&\Pi^{\theta\sigma}(0,0)\equiv
-i{\mathcal{FT}}\{\langle0|\theta\sigma|0\rangle\},
\end{eqnarray}where ${\mathcal{FT}}\{\cdots\}$ denotes the Fourier
transform and $m$ refers to the fermion mass. The canonical
identities for trace relation and chiral symmetry that should be
satisfied by the above vertex functions\cite{PCDC} are as follows,
\begin{eqnarray}
\label{traceid1}
&&\Delta^5_{\mu\nu}(0,p,-p)=(2-p\partial_p){\Pi}^5_{\mu\nu}(p,-p),\\
\label{traceid2} &&\Delta^5_{\
\nu}(0,p,-p)=(2-p\partial_p){\Pi}^5_{\ \nu}(p,-p),\\
\label{traceid3}&&\Delta^5(0,p,-p)=(2-p\partial_p){\Pi}^5(p,-p);\\
\label{chiralid1} &&-ip^{\mu}\Delta^5_{\mu\nu}(0,p,-p)=\Delta^5_{\
\nu}(0,p,-p)+{\Pi}^5_{\ \nu}(p,-p),\\
\label{chiralid2}&&ip^{\nu}\Delta^5_{\
\nu}(0,p,-p)=\Delta^5(0,p,-p) +{\Pi}^5(p,-p)+
\Pi^{\theta\sigma}(0,0),\\
\label{chiraldi3}&&-ip^{\mu}{\Pi}^5_{\mu\nu}(p,-p)={\Pi}^5_{\
\nu}(p,-p),\\
\label{chiraldi4}&&ip^{\nu}{\Pi}^5_{\ \nu}(p,-p)={\Pi}^5(p,-p)
+\langle\sigma\rangle.
\end{eqnarray}The first three are canonical trace identities and the rest
are canonical chiral Ward identities.

After some calculations in dimensional regularization we obtain
the following one-loop results for the interested objects:
\begin{eqnarray}
\label{1-loop}
{\Pi}^{5}_{\mu\nu}(p,-p)&=&\frac{2g_{\mu\nu}}{(4\pi)^2} \{
p^2[\Delta_0-\Gamma(\epsilon)+2\int^1_0 dx (x^2-x)(\ln
\frac{D}{4\pi\mu^2}-\Gamma(\epsilon)-1)]\nonumber \\
&&-2m^2[\Delta_0+\ln\frac{m^2}{4\pi\mu^2}-2\Gamma(\epsilon)]\}\nonumber \\
&&+\frac{p_{\mu}p_{\nu}}{4\pi^2}\{2\int^1_0 dx (1-x)^2[\ln
 \frac{D}{4\pi\mu^2}-\Gamma(\epsilon)]+\Gamma(\epsilon)-\Delta_0\},\\
\Delta^{5}_{\mu\nu}(0,p,-p)&=&\frac{g_{\mu\nu}m^2}{2\pi^2}
\{2\Gamma(\epsilon)-\ln\frac{m^2}{4\pi\mu^2}-\Delta_0
+\frac{p^2/2-2m^2}{\Delta}
\}+\frac{m^2p_{\mu}p_{\nu}}{\pi^2p^2}(1-\frac{m^2}{\Delta});\\
{\Pi}^{5}_{\
\nu}(p,-p)&=&\frac{im^2p_{\nu}}{2\pi^2}(\Delta_0-\Gamma(\epsilon)),\\
\Delta^{5,\epsilon}_{\ \nu}(0,p,-p)
&=&\frac{im^2p_{\nu}}{2\pi^2}(\Delta_0-\Gamma(\epsilon)+\frac{2m^2}{\Delta});\\
{\Pi}^{5}(p,-p)&=&\frac{m^2}{2\pi^2}
\{2m^2(\ln\frac{m^2}{4\pi\mu^2}-\Gamma(\epsilon)-1)
-p^2(\Delta_0-\Gamma(\epsilon))
\},\\
\Delta^{5}(0,p,-p)&=&\frac{m^4}{\pi^2}
\{2(\ln\frac{m^2}{4\pi\mu^2}-\Gamma(\epsilon))-
\frac{p^2}{\Delta}\};\\
\langle\sigma\rangle&=&\frac{m^4}{\pi^2}
(\Gamma(\epsilon)+1-\ln\frac{m^2}{4\pi\mu^2}),\\
\Pi^{\theta\sigma}(0,0)&=&\frac{3m^4}{\pi^2}
(\Gamma(\epsilon)+1/3-\ln\frac{m^2}{4\pi\mu^2}),
\end{eqnarray}with $
D=m^2+p^2(x^2-x), \Delta_0=\int^1_0 dx \ln \frac{D}{4\pi\mu^2},
\frac{1}{\Delta}=\int^1_0 \frac{dx}{D}$. Now we could check them
with our results given above. Inserting these functions into Eqs.
(~\ref{traceid1}, ~\ref{traceid2}, ~\ref{traceid3},
~\ref{chiralid1}, ~\ref{chiralid2}, ~\ref{chiraldi3},
~\ref{chiraldi4}), we find that all the chiral identities are
valid for the one-loop functions calculated above in dimensional
regularization, but all the trace identities are violated, namely,
Eq.s(~\ref{traceid1},~\ref{traceid2},~\ref{traceid3}) are modified
as follows,
\begin{eqnarray}
\label{atraceid1}
&&\Delta^5_{\mu\nu}(0,p,-p)=(2-p\partial_p){\Pi}^5_{\mu\nu}(p,-p)
+\frac{1}{6\pi^2}(g_{\mu\nu}p^2-p_{\mu}p_{\nu})-
\frac{m^2}{\pi^2}g_{\mu\nu},\\
\label{atraceid2} &&\Delta^5_{\
\nu}(0,p,-p)=(2-p\partial_p){\Pi}^5_{\ \nu}(p,-p)+
\frac{im^2}{\pi^2}p_{\nu},\\
\label{atraceid3}&&\Delta^5(0,p,-p)=(2-p\partial_p){\Pi}^5(p,-p)
-\frac{m^2}{\pi^2}p^2+\frac{2m^4}{\pi^2}
\end{eqnarray}
Now anomalies appear in all the three trace identities. That means
in dimensional regularization, the chiral Ward identities are
preserved in these three- and two-point functions for the axial
operators considered, but the trace identities are quantum
mechanically violated. The above results are obtained without use
of the chiral Ward identities, unlike the procedures taken in
Ref.\cite{PCDC}. To compare our results with previous ones and to
check if the chiral Ward identities are consistent with these
anomalous trace identities, we follow the procedures of
Ref.\cite{PCDC}. We should also note that these anomalous
identities still hold even after minimal-like subtraction, that
is, they are valid for both unrenormalized and renormalized vertex
functions.

That is, we apply the relations
Eq.s(~\ref{chiralid1},~\ref{chiralid2},~\ref{chiraldi3},~\ref{chiraldi4})
to the first of the anomalous equation to derive the other two.
Noting that
\begin{equation}
\label{sigma} \frac{2m^4}{\pi^2}=
3\langle\sigma\rangle-\Pi^{\theta\sigma}(0,0),
\end{equation}
we arrive at the following form of the anomalous trace identities:
\begin{eqnarray}
\label{atraceid10}
&&\Delta^5_{\mu\nu}(0,p,-p)=(2-p\partial_p){\Pi}^5_{\mu\nu}(p,-p)
+\frac{1}{6\pi^2}(g_{\mu\nu}p^2-p_{\mu}p_{\nu})-
\frac{m^2}{\pi^2}g_{\mu\nu},\\
\label{atraceid20} &&\Delta^5_{\
\nu}(0,p,-p)=(2-p\partial_p){\Pi}^5_{\ \nu}(p,-p)+
\frac{im^2}{\pi^2}p_{\nu},\\
\label{atraceid30}&&\Delta^5(0,p,-p)=(2-p\partial_p){\Pi}^5(p,-p)
-\frac{m^2}{\pi^2}p^2+3\langle\sigma\rangle-\Pi^{\theta\sigma}(0,0).
\end{eqnarray}Now we find complete agreement between the two
approaches, since
Eqs.(~\ref{atraceid10},~\ref{atraceid20},~\ref{atraceid30}) are
exactly the same as
Eqs.(~\ref{atraceid1},~\ref{atraceid2},~\ref{atraceid3}), due to
the relation given Eq.(~\ref{sigma}).

However, comparing
Eqs.(~\ref{atraceid10},~\ref{atraceid20},~\ref{atraceid30}) or
Eqs.(~\ref{atraceid1},~\ref{atraceid2},~\ref{atraceid3}) with
those in Ref.\cite{PCDC}, we find two disagreements: (1) In
Ref.\cite{PCDC}, the numerical coefficient of the anomaly term
$(g_{\mu\nu}p^2-p_{\mu}p_{\nu})$ is $\frac{1}{8\pi^2}$ while here
in Eq.(~\ref{atraceid10}) it is $\frac{1}{6\pi^2}$; (2) In
Ref.\cite{PCDC} the last two terms in Eq.(~\ref{atraceid30}) (or
the $\sim m^4$ term in Eq.(~\ref{atraceid3})) were missing.

The most interesting anomalous identity is Eq.(~\ref{atraceid30}).
From the trace identity perspective, both $-\frac{m^2}{\pi^2}p^2$
and $3\langle\sigma\rangle-\Pi^{\theta\sigma}(0,0)$ are anomalies.
However, the latter is required by and explicable within the
chiral Ward identities and its existence is independent of
regularization or short distance physics, thus we find an
interesting phenomenon: {\em the canonical terms in chiral
identity become anomalies in trace identity}. To our knowledge,
this phenomenon has not yet been reported in field theory and high
energy physics literature. It is known that in supersymmetric
field theories, the gauge field components from the trace
anomalies ($\sim \text{tr}(F^2)$) and from chiral anomalies ($\sim
\text{tr}(F\widetilde{F})$) comprise a supermultiplet and hence
share the same coefficient\cite{susyanomaly}, thus chiral
'symmetry' and scale 'symmetry' are closely related in
supersymmetric contexts. Here we encountered another phenomenon
between the trace identities and chiral identities, in the axial
scalar sectors as the 'current' density $j_5$ couples to axial
scalar fields. The deeper implications of this interesting finding
is not clear for us yet. We refrain from making speculations about
it before further investigation is carried out. Whilst we believe
that it is worthwhile to pay attention to this phenomenon. Lastly
we mention that this phenomenon is independent of regularization
scheme, for we have also calculated all the vertex functions in a
general parametrization of regularization schemes and reobtained
Eq.(~\ref{atraceid30}), for details see\cite{311}.

In summary, we have investigated the trace identities and chiral
identities for certain vertex functions of axial operators, by one
loop calculations in dimensional regularization directly. Some
disagreements with previous publications and an interesting
phenomenon were found.
\section*{Acknowledgement} The author is grateful to W. Zhu for
his continuing supports and helpful conversations on the topics
related to scaling. This work is supported in part by the National
Natural Science Foundation of China under Grant No.s 10075020 and
10205004.

\end{document}